\documentclass[aps,prd,superscriptaddress,twoside,twocolumn,nofootinbib,10pt,
showpacs,floatfix]{revtex4-1}
\usepackage{amsmath,amssymb}
\usepackage[dvipdfmx]{graphicx}
\usepackage[dvipdfmx]{}
\usepackage{bm}
\usepackage{overpic}
\usepackage{ulem}
\usepackage[usenames]{color}
\usepackage{float}
\usepackage{color}
\usepackage{braket}

\def\slash#1{\not\!#1}

\allowdisplaybreaks

\begin{document}

\author{Yusuke Takeda}
\affiliation{Department of Physics, Nagoya University, Nagoya, 464-8602, Japan}
\author{Hiroaki Abuki}
\affiliation{Department of Education, Aichi University of Education,
Kariya, 448-8542, Japan}

\author{Masayasu Harada}
\affiliation{Department of Physics, Nagoya University, Nagoya, 464-8602, Japan}

\title{A novel Dual Chiral Density Wave in nuclear matter\\ based on a parity doublet structure}

\begin{abstract}

We study the Dual Chiral Density Wave (DCDW) in nuclear matter using a
 hadronic model with the parity doublet structure.
We first extend the ordinary DCDW ansatz so as to incorporate the
 effect of an explicit chiral symmetry breaking.
Then via numerically evaluating and minimizing the effective potential,
 we determine the phase structure. 
We find, in addition to the ordinary DCDW phase where the space average
 of the chiral condensate vanishes, a new DCDW phase (sDCDW)
with a nonvanishing space average 
depending on the value of the chiral
 invariant mass parameter.

\end{abstract}

\pacs{}

\maketitle

\section{Introduction}
Last few decades, 
the phase structure of QCD has 
been one of the main concerns regarding the physics of strong
interaction.
For the baryon chemical potential lower than the nucleon mass minus
binding energy per baryon in nuclei, the phase realized in  nature is
the QCD vacuum where the chiral symmetry is spontaneously broken.
The chiral symmetry breaking is responsible for the mass generation of
hadrons as well as the mass splittings of chiral partners.

At sufficiently large chemical potential and/or temperature, the chiral
symmetry is expected to restore.
An interesting possibility arises when one allows the chiral condensate
to vary in space; %
Nakano and Tatsumi demonstrated  in \cite{Nakano:2004cd} using the NJL
model that the symmetry restoration may take place via several steps;
going up in density from the vacuum, the system first goes into an
intriguing
state named as dual chiral density wave (DCDW), that is, 
a particular type of inhomogeneous chiral phases, making a spiral in the
$(\sigma_0,\pi_0)$ chiral plane along $z$ direction.
Up to present, various inhomogeneous chiral phases are discussed.
These includes the real kink crystal (RKC) phase belonging to another
class of inhomogeneous phases \cite{Nickel:2009ke}.
In both cases, the chiral symmetry is partially restored in either the
momentum space (DCDW) or the real space (RKC).
Such inhomogeneous chiral phases may be realized in the neutron
stars and may lead to some interesting astrophysical implications
\cite{Tatsumi:2014cea,Buballa:2015awa}.

There are number of approaches to chiral inhomogeneous phases.
One of the major strategies is to apply the mean-field approximation
\cite{Nickel:2009wj,Carignano:2010ac,Karasawa:2013zsa,Adhikari:2017ydi}
or the Ginzburg-Landau (gradient) expansion
\cite{Abuki:2011pf,Abuki:2013pla,Carignano:2017meb} 
to the quark-based models such as NJL-type models or quark-meson model
\cite{Buballa:2014tba}.
Recently a self-consistent mean-field framework has also been applied
\cite{Lee:2017yea}.
One of the advantages of 
this kind of approach
is that these models are capable of realizing the QCD vacuum properties
as well as the color-flavor locked phase of quark matter which is known
to be the densest phase of QCD \cite{Alford:2007xm}.
On the other hand, the main disadvantage is the lack of the ability to
reproduce the normal nuclear matter, the QCD phase next to the vacuum
phase, followed right after the liquid-gas phase transition.
A quite different approach was taken recently in \cite{Heinz:2013hza}.
Using the parity doublet hadron model tuned to reproduce the bulk
properties of normal nuclear matter, the authors have shown that the
DCDW phase appears at density several times larger than the normal
nuclear density.

In the present paper, we adopt a hadronic model with parity doublet
structure (mirror assignment) \cite{Detar:1988kn,Jido:2001nt}, with
vector mesons included in a manner guided by the hidden local symmetry
\cite{Bando:1987br,Harada:2003jx}.
With the six-point scalar interaction included, the model is known to
successfully reproduce bulk properties of normal nuclear matter for a
wide range of chiral invariant mass \cite{Motohiro:2015}.
Our main concerns here are, 1) if the inhomogeneous chiral phase is
possible or not within our model, 2) how phase transition points, if
any, as a function of $\mu_B$, change with the chiral invariant mass,
and 3) what is the effect of current quark mass on the inhomogeneous
chiral phase.
In particular, the point 3) was missed in \cite{Heinz:2013hza}.
In order to incorporate this in our analyses, we extend the ansatz for
the DCDW phase so as to take into account the effect of the explicit
symmetry breaking. 
The extended ansatz smoothly interpolates between the DCDW phase and a
nearly symmetry-restored phase.
With this setup, we construct the effective potential by diagonalizing
the 
Bogoliubov-de Gennes (BdG) Hamiltonian for nucleons, and determine phases
via numerically minimizing the potential.
Our main finding is the emergence of another type of DCDW phase which
occupies the lower density region according to the value of chiral
invariant mass.

The paper is organized as follows.
In Sec.~\ref{sec:model}, we describe our model setup and approximation
scheme.
In Sec.~\ref{sec:PhaseStructure}, we present our numerical results for
phases and discuss the phase structure in the plane of $\mu_B$ and
chiral invariant mass.
Sec.~\ref{sec:summary} summarizes the present work.

\section{Model}\label{sec:model}
In our analysis, we introduce $N^\ast(1535)$ as the chiral partner to
the ordinary nucleon based on the parity doublet
structure~\cite{Detar:1988kn,Jido:2001nt}.  
For constructing a relativistic mean field model to describe nuclear
matter, following Ref.~\cite{Motohiro:2015}, we include $\omega$ meson
using the hidden local symmetry~\cite{Bando:1987br,Harada:2003jx} in
addition to the scalar and pseudoscalar mesons.
The baryon part of the Lagrangian  is expressed as~\cite{Motohiro:2015}
\begin{align}
 {\cal L}_N=&\bar\psi_{1r}i\gamma^\mu D_\mu\psi_{1r}%
            +\bar\psi_{1l}i\gamma^\mu D_\mu\psi_{1l}&\nonumber \\
           &+\bar\psi_{2r}i\gamma^\mu D_\mu\psi_{2r}%
            +\bar\psi_{2l}i\gamma^\mu D_\mu\psi_{2l}&\nonumber \\
           &-m_0[\bar\psi_{1l}\psi_{2r}-\bar\psi_{1r}\psi_{2l}%
            -\bar\psi_{2l}\psi_{1r}+\bar\psi_{2r}\psi_{1l}]&\nonumber\\
           &-g_1[\bar\psi_{1r}M^\dagger\psi_{1l}%
            +\bar\psi_{1l}M\psi_{1r}]&\nonumber\\
	   &-g_2[\bar\psi_{2r}M\psi_{2l}%
            +\bar\psi_{1l}M^\dagger\psi_{2r}]&\\
	   &+a_{\rho NN}[\bar\psi_{1l}\gamma^\mu\xi_L^\dagger%
            \hat\alpha_{\parallel\mu}\xi_L\psi_{1l}
            +\bar\psi_{1r}\gamma^\mu\xi_R^\dagger%
            \hat\alpha_{\parallel\mu}\xi_R\psi_{1r}]&\nonumber\\
           &+a_{\rho NN}[\bar\psi_{2l}\gamma^\mu\xi_R^\dagger%
            \hat\alpha_{\parallel\mu}\xi_R\psi_{2l}%
            +\bar\psi_{2r}\gamma^\mu\xi_L^\dagger%
            \hat\alpha_{\parallel\mu}\xi_L\psi_{2r}]&\nonumber\\
           &+a_{0NN}{\rm tr}[\hat\alpha_{\parallel\mu}]%
            (\bar\psi_{1r}\gamma^\mu\psi_{1r}+\bar\psi_{1l}%
            \gamma^\mu \psi_{1l}&\nonumber\\
           & \ \ \ \ \ \ \ \ \ \ \ \ \ \ \ \ \ \ \ \ \ \ \ \ \  %
            +\bar\psi_{2r}\gamma^\mu \psi_{2r}%
            +\bar\psi_{2l}\gamma^\mu \psi_{2l})&\nonumber
\label{eq:LN}
\end{align}
The part for the scalar and pseudoscalar mesons meson part is given by
\begin{align}
 {\cal L}_M=&\frac{1}{4}{\rm tr}\left[%
            \partial_\mu M\partial^\mu M^\dagger%
            \right]+\frac{1}{4}\bar\mu^2{\rm tr}\left[%
            MM^\dagger\right]&\nonumber \\
           &-\frac{1}{16}\lambda_4\left({\rm tr}\left[%
             MM^\dagger\right]\right)^2+\frac{1}{48}%
             \lambda_6\left({\rm tr}\left[%
             MM^\dagger\right]\right)^3&\nonumber \\
	   &+\frac{1}{4}m_\pi^2 f_\pi{\rm tr}\left[%
             M+M^\dagger\right]&
\end{align}
In this paper, we omit the kinetic and mass terms for vector mesons.
Details of the above Lagrangian terms are seen in \cite{Motohiro:2015}.

In the present analysis, we adopt the following extended DCDW ansatz
\begin{align}
 \braket{M}=M(z)\equiv\delta\sigma+\sigma_0e^{2ifz\tau^3}
\label{eq:ansatz}
\end{align}
where $\delta \sigma$, $\sigma_0$ and $f$ are parameters with dimension
one, and $\tau^a$ ($a=1,2,3$) are the Pauli matrices.
Space independent part $\delta\sigma$ accommodates the possibility that
the space average of DCDW condensate would get nonvanishing shift into
$\sigma$-direction due to the explicit chiral symmetry breaking.
Applying the mean-field approximation, the Lagrangian for nucleon is
cast into 
\begin{align}
{\cal L}_N=&\bar\psi_1\left[i\slash{\partial}-g_1\left(%
            \delta\sigma+\sigma_0e^{2ifz\tau^3\gamma_5}%
            \right)+\gamma^0\mu_B^*\right]\psi_1&\nonumber \\
	   &+\bar\psi_2\left[i\slash{\partial}-g_2\left(%
            \delta\sigma+\sigma_0e^{-2ifz\tau^3\gamma_5}%
            \right)+\gamma^0\mu_B^*\right]\psi_2&\nonumber \\
	   &-m_{0}\left(\bar\psi_1\gamma_5\psi_2%
            -\bar\psi_2\gamma_5\psi_1\right)&
\end{align}
where $\mu_B^*$ is effective chemical potential which include $\omega$
contribution as
\begin{equation*}
	\mu_B^*=\mu_B-g_{\omega NN}\omega_0\, .
\end{equation*}
The nucleon contribution to the effective potential can be written as
\begin{align}
 \Omega_{N}=\frac{i}{V_4}\mathrm{Tr}\,%
 \mathrm{Log}(i\partial_0-({\mathcal H}(z)-\mu_B^*)),
\label{eq:omegaB}
\end{align}
where $V_4$ is the space-time volume and ${\mathcal H}(z)$ is the single
particle Bogoliubov-de Gennes (BdG) Hamiltonian defined in the space of
fermion bispinor $\psi=(\psi_1,\psi_2)$ as
\begin{equation*}
{\mathcal H}=\left(%
\begin{array}{cc}
i\gamma^0\bm{\gamma}\cdot\nabla+g_1\gamma^0M(z) & m_{0}\gamma^0\gamma_5 \\
-m_{0}\gamma^0\gamma_5 & i\gamma^0\bm{\gamma}\cdot\nabla+g_2\gamma^0M(z)^* \\
\end{array}
\right)
\end{equation*} 
This is nothing but the Dirac Hamiltonian in the presence of a periodic
potential field $M(z)(=M(z+\frac{\pi}{f}))$. 
Then the functional trace in Eq.~(\ref{eq:omegaB}) can be evaluated by
finding eigenvalues of the operator ${\mathcal H}(z)$
\cite{Nickel:2008ng}.
The eigenvalue has a discrete label as well as continuous three-momentum
${\bf p}$ in addition to internal quantum numbers; 
This is because of the Bloch theorem which states that the
eigenfunctions in the presence of a periodic potential are the modified
plane 
wave; the plane 
wave distorted by periodic functions.
To be specific, we decompose the bispinor as
\begin{equation*}
\psi({\bf x})=\sum_{\ell=-\infty}^{\infty}\sum_{\bf p}%
\psi_{{\bf p},\ell}\,e^{i\left({\bm K}_\ell+{\bf p}\right)\cdot{\bf x}}
\end{equation*}
where ${\bm K}_\ell=(0,0,2f\ell)$ is the reciprocal lattice vector.
Moving on to the quasimomentum base $\{\psi_{{\bf p},\ell}\}$, the BdG
Hamiltonian for proton $(I_3=+1/2)$ sector is cast into the following
block-diagonalized form:
\begin{align}
 H_{\ell\ell'}({\bf p})=%
  &\left( %
     \begin{array}{@{\,}cccc@{\,}}
      H^{1}_{\ell\ell'}&\gamma^0\gamma_5m_0\delta_{\ell\ell'}\\
    -\gamma^0\gamma_5m_0\delta_{\ell\ell'}&H^{2}_{\ell\ell'}
     \end{array}%
    \right)&\nonumber \\
 H^{1}_{\ell\ell'}=%
  &\left[\left({\bf p}+{\bm K}_\ell\right)\cdot\gamma^0%
   {\bm\gamma}+g_1\delta\sigma\gamma^0\right]\delta_{\ell\ell'}%
  &\nonumber \\
  &+g_1\sigma_0\gamma^0\left[P_r\delta_{\ell\ell'+1}%
   +P_l\delta_{\ell\ell'-1}\right]%
  &\nonumber \\
    H^{2}_{\ell\ell'}=&\left[\left({\bf p}+{\bm K}_\ell\right)%
    \cdot{\gamma^0\bm \gamma}+g_2\delta\sigma\gamma^0\right]%
    \delta_{\ell\ell'}&\nonumber \\
  &+g_2\sigma_0\gamma^0\left[P_r\delta_{\ell\ell'-1}%
   +P_l\delta_{\ell\ell'+1}\right]%
  &\nonumber 
\end{align}
where $P_{r,l}$ is projection operator defined as
\begin{equation*}
 P_{r}=\frac{1+\gamma_5}{2} \ \ , \ \ P_{l}=\frac{1-\gamma_5}{2}.
\end{equation*}
Since the isospin remains a good quantum number, we can simply double
the proton contribution in the full effective potential.
Then omitting the antiprotons which would not contribute at zero
temperature, and diagonalizing $H_{\ell\ell^\prime}({\bf p})$ results in
an infinite tower of eigenvalues at each ${\bf p}$, which 
repeatedly appears for every Brillouin %
Zone (BZ), ${\bf p}\to {\bf p}+\bm{K}_\ell$ ($\ell=\cdots,-1,0,1,\cdots$):
\begin{equation*}
\sum_{\ell^\prime}%
 H_{\ell\ell'}({\bf p})\psi^{(i)}_{n,{\bf p},\ell^\prime}%
 =E_{n,\bf{p}}^{(i)}\psi^{(i)}_{n,\bf{p},\ell}\,%
 \quad(n=0,1,\cdots,\infty),
\end{equation*}
with $i(=1, 2, 3, 4)$ labeling the internal quantum number
$(p,p^*)\otimes(\uparrow,\downarrow)$, where $p^\ast$ implies the
$I_3=+1/2$ part of $N^\ast(1535)$.
Equation~(\ref{eq:omegaB}) is now evaluated as
\begin{align}
\Omega_N=\sum_{n=0}^{\infty}%
 \sum_{i=1}^4\int_{-f}^f\frac{dp_z}{\pi}%
 \!\!\int\frac{d{\bf p}_\perp}{(2\pi)^2}%
 (E^{(i)}_{n,{\bf p}}-\mu_B^*)\theta(\mu_B^*-E^{(i)}_{n,{\bf p}})
\label{eq:OmegaN}
\end{align}
with ${\bf p}_\perp=(p_x,p_y,0)$.
The following meson contributions add up to the full expression of the
thermodynamic potential.
\begin{align}
 \Omega_M=&-\frac{1}{2}m_\omega^2\omega_0^2%
           -\frac{1}{2}\bar\mu^2\left(%
           \delta\sigma^2+\sigma_0^2%
           \right)+2\sigma_0^2f^2&\nonumber \\
          &+\frac{1}{4}\lambda_4\left[%
           \left(\delta\sigma^2+\sigma_0^2\right)^2%
           +2\delta\sigma^2\sigma_0^2%
           \right]&\nonumber \\
          &-\frac{1}{6}\lambda_6\left[%
           \left(\delta\sigma^2+\sigma_0^2\right)^3%
           +6\left(\delta\sigma^2+\sigma_0^2\right)%
           \delta\sigma^2\sigma_0^2\right]&\nonumber \\
          &-m_\pi^2f_\pi\delta\sigma.&
\label{eq:Omega}
\end{align}
The feedback from an explicit chiral symmetry breaking is taken care
by the last term.

Assuming for a while the existence of normal nuclear matter within the
model, model parameters except for chiral invariant mass for nucleon are
determined by the pion decay constant $\sigma_0=f_\pi=92.2$~MeV  in
vacuum, baryon and meson masses shown in Table \ref{table:input-mass}
and normal nuclear property shown in
Table~\ref{table:input-normal-nuclear-density}.
In homogeneous phase the baryon mass is calculated as
\begin{align}
 m_{\pm}=\frac{1}{2}\left[%
   \sqrt{\left(g_1+g_2\right)^2\sigma_0^2+4m_0^2}%
   \mp\left(g_1-g_2\right)\sigma_0%
   \right].
\end{align}
The determined parameters are summarized in Table
\ref{table:model-para}.
However, we will show later that the normal nuclear matter exists only
as a metastable state as another type of DCDW phase dominates over it
once chiral invariant mass becomes smaller than some critical value,
$m_{0}\alt 800$~MeV.
\begin{table}[thbp]
\caption{Values of baryon and meson mass in unit of MeV }
\label{table:data_type}
\centering
\begin{tabular}{cccc} 
\hline \hline
$m_+$&$m_- $&$m_\pi$&$m_\omega$\\
\ \ 939\ \ &\ \ 1535\ \ &\ \ 140\ \ &\ \ 783\ \ \\
\hline\hline
\end{tabular}
\label{table:input-mass}
\end{table}
\begin{table}[thbp]
\caption{ Physical inputs in normal nuclear density}
\label{table:input-normal-nuclear-density}
\centering
\begin{tabular}{cc|c} 
\hline\hline
\ \ Saturation density\ \   & \ \ $\rho_0$ \ \ &%
 \ \ 0.16 [${\rm fm}^{-3}$] \ \ \\
\ \ Binding energy \ \ & \ \ $\frac{E}{A}-939$ \ \ &%
\ \ $-16$ [MeV] \ \ \\
\ \ Incompressibility \ \ &$ \ \ K \ \ $& \ \ 240 [MeV] \ \ \\
\hline\hline
\end{tabular}
\end{table}
\begin{table}[h]
\caption{Determined parameters for given chiral invariant mass}
\label{table:model-para}
\centering
\begin{tabular}{c|cccccc}
\hline\hline
\ $m_{0}$ \ & \ $500$ \ & \ $600$ \ & \ $700$ \ & \ $800$ \ & 900 \ \\
\hline
$g_1$&$9.03$&8.49&7.82&7.00&5.97\\
$g_2$&$15.5$&15.0&14.3&13.5&12.4\\
$g_{\omega NN}$&11.3&9.13&7.30&5.66&3.52\\
$\bar\mu\,[\rm{MeV}]$&441&437&406&320&114\\
$\lambda_4$&42.2&40.6&35.7&23.2&4.47\\
$\lambda_6\cdot f_\pi^2$&17.0&15.8&14.0&8.94&0.644\\
\hline\hline
\end{tabular}
\end{table}


\section{phase structure}\label{sec:PhaseStructure}
The phase diagram can be obtained by numerically solving the stationary
conditions of thermodynamic potential for $f$, $\delta\sigma$,
$\sigma_0$ and $\omega_0$. 

\begin{figure}[tbp]
\begin{center}
\vspace{-80mm}
\includegraphics[bb=0 0 480 600,width=10cm,clip]{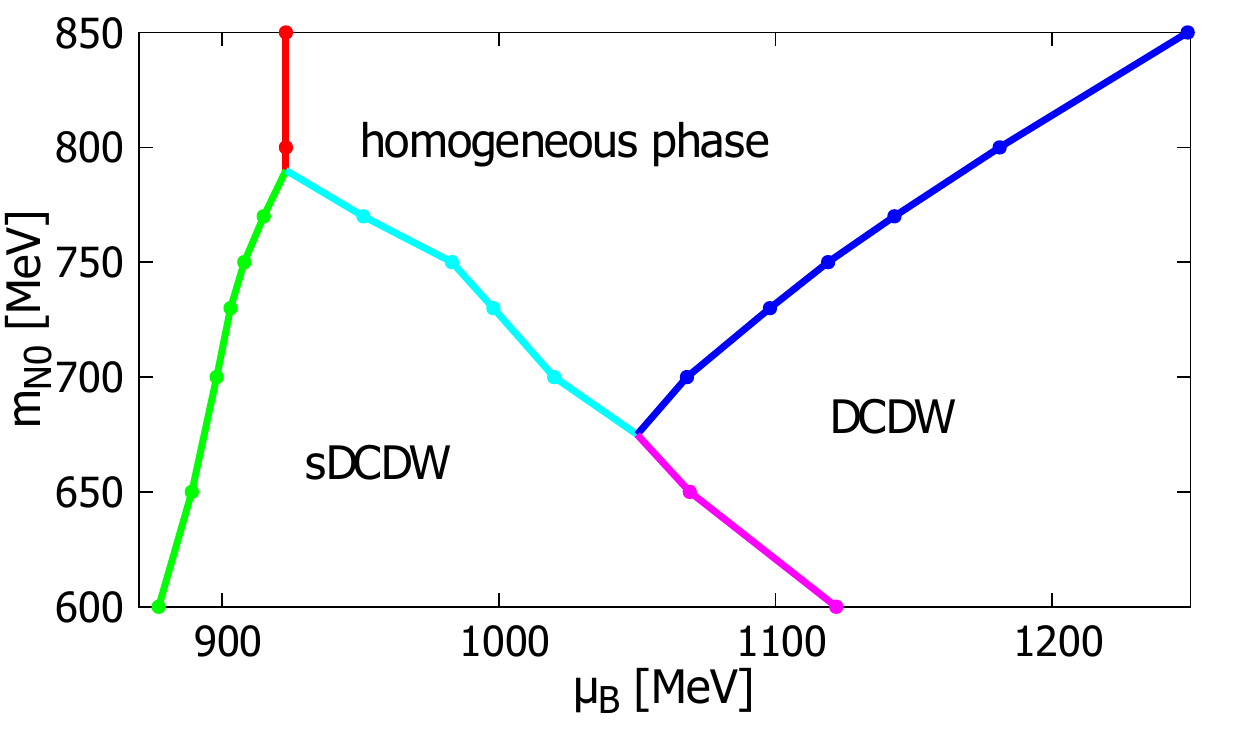}
\caption{Phase structure in $\mu_B-m_{0}$ plane. }
\label{fig:muB-mN0}
\end{center}
\end{figure}

In Figure~\ref{fig:muB-mN0}, we show the phase structure in
$\mu_B-m_{0}$ plane.
We find that there exist two kinds of DCDW phase: the ordinary DCDW
phase indicated as ``DCDW'' and a new DCDW phase as ``sDCDW''.

In the following, we discuss the phases and the associated phase
transitions in detail for two typical cases (a)~$m_{0}=800$MeV, and
(b)~$m_{0}=700$MeV. 

\begin{figure}[tbp]
\begin{center}
\includegraphics[width=7cm,clip]{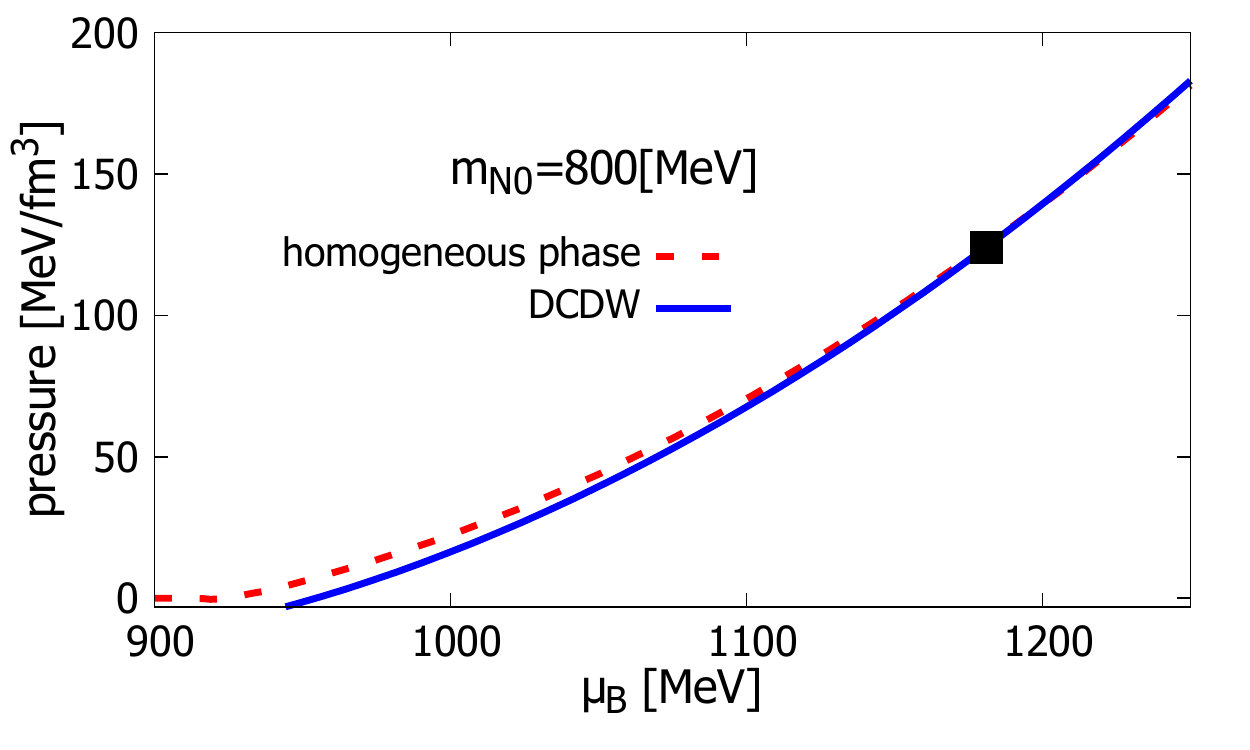}
\end{center}
\caption{
Relation between chemical potential and pressure for $m_{0}=800$~MeV.
The blue solid curve and red dashed curve show DCDW phase and
 homogeneous phase, respectively.
The point of phase transition from the homogeneous phase to the DCDW
 phase is expressed by black square.
}
\label{fig:muB-pres-800-a}
\end{figure}

\paragraph{$m_{0}=800$MeV.}
Figure~\ref{fig:muB-pres-800-a} shows the pressure for the homogeneous
($f=0$, a dashed curve) and DCDW states ($f\ne0$, a solid curve) as a
function of $\mu_B$ in this case.
\begin{figure}[tbp]
\begin{center}
\includegraphics[width=7cm,clip]{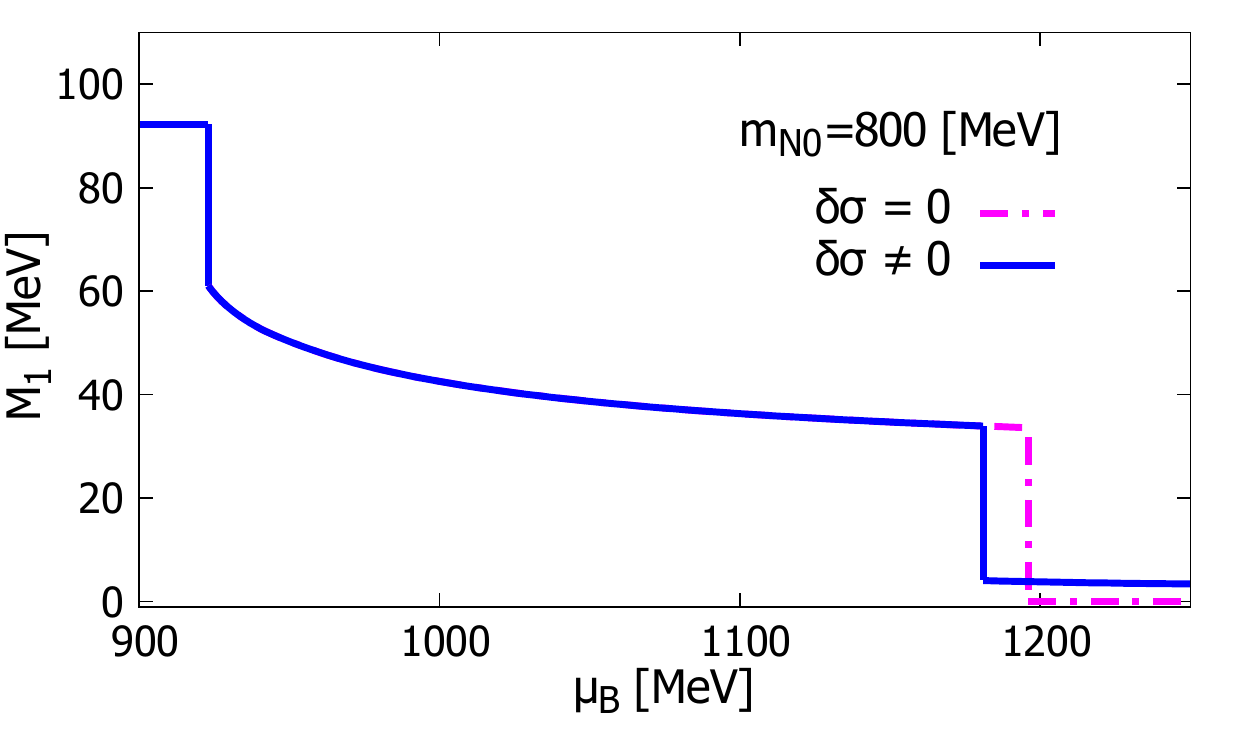}
\end{center}
\caption{
Relation between chemical potential and $M_1$ for $m_{0}=800$~MeV.
The blue solid curve and magenta dot-dashed curve show the configuration
 of the extended ansatz with $\delta\sigma \neq0$ and the ordinary
 ansatz with $\delta\sigma = 0$, respectively. 
}
\label{fig:muB-pres-800-b}
\end{figure}
These two states consist only two stationary solutions for
$\delta\Omega=0$ for a given value of $\mu_B$.
The DCDW state found here corresponds to the ordinary DCDW state
in nuclear matter \cite{Heinz:2013hza}, but here the chiral
condensate $M(z)$ has a finite offset $\delta\sigma$ due to an explicit
symmetry breaking.
The ordinary DCDW state takes over the homogeneous nuclear matter at
$\mu_B\sim1200$~MeV denoted in the figure by square.

In what follows, for the comparison with the ordinary ansatz, we define
the thermodynamic potential with the condition $\delta\sigma=0$ is
forced by
\begin{equation*}
\Omega_0=\Omega\mid_{\delta\sigma=0}.
\end{equation*}
Figure~\ref{fig:muB-pres-800-b} shows one of order parameter $M_1$
defined by
\begin{align}
 M_1\equiv&\frac{1}{2V}\int_{-\infty}^{\infty}d^3x%
           {\rm tr}\left[\braket{M}\right]&\nonumber \\
         =&\begin{cases}
	    \delta\sigma+\sigma_0 & (f=0) \\
	    \delta\sigma & (f\neq0)\\
	   \end{cases}
 \label{eq:M1}
\end{align}
This quantity may be regarded roughly as a guide for the strength of
chiral symmetry breaking.
We see that solid curve in the region of the DCDW phase is nonvanishing,
indicating that the effect of explicit symmetry breaking is properly
taken into account in our extended ansatz.

\begin{figure}[tbp]
\begin{center}
\includegraphics[width=7cm,clip]{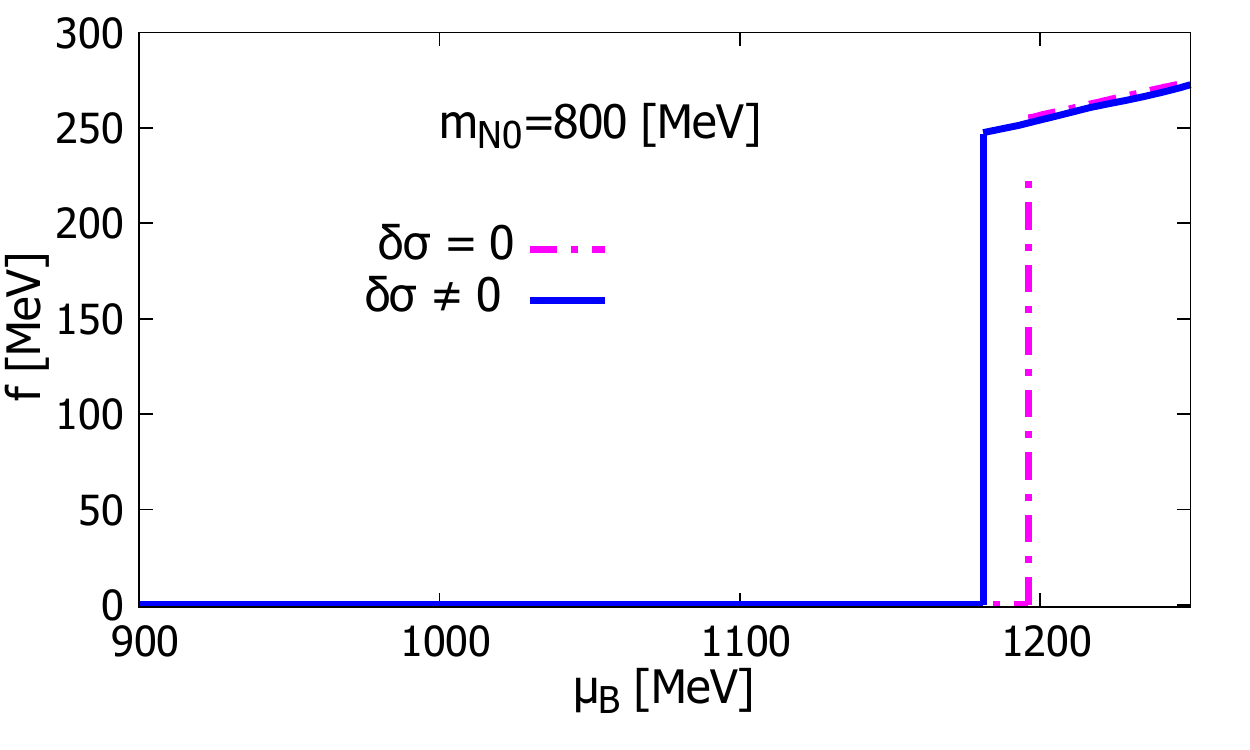}
\end{center}
\caption{
Relation between chemical potential and $f$ for $m_{0}=800$~MeV.
The blue solid curve and magenta dot-dashed curve show the configuration
 of the extended ansatz with $\delta\sigma \neq 0$ and the ordinary
 ansatz with $\delta \sigma =0$, respectively.
 }
\label{fig:muB-pres-800-c}
\end{figure}

\begin{figure}[tbp]
\begin{center}
\includegraphics[width=7.cm,clip]{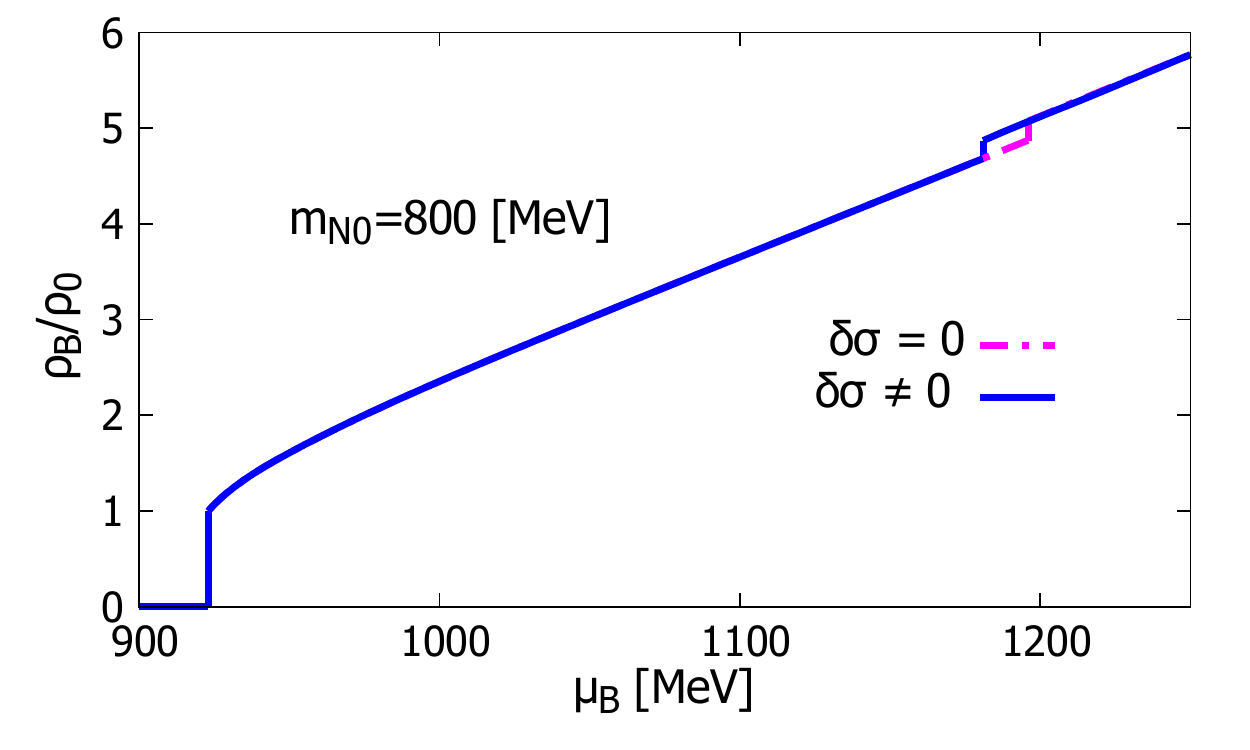}
\caption{
 Relation between chemical potential and $f$ for $m_{0}=800$~MeV.
 The blue solid curve and magenta dot-dashed curve show the
 configuration of the extended ansatz with $\delta \sigma \neq0$ and
 the ordinary ansatz with $\delta \sigma=0$, respectively.
 }
 \label{fig:muB-density-800}
\end{center}
\end{figure}

Depicted in Figure~\ref{fig:muB-pres-800-c} is the wave number $f$ as a
function of $\mu_B$. 
The quantity provides a guide for a translational symmetry breaking.
We note that, even though the effect of $\delta\sigma$ gives a minor
effect on the magnitude of $f$, it brings about a sizable shift of the
critical chemical potential, making the onset point several tens of MeV
earlier than the case of ordinary ansatz.
Figure~\ref{fig:muB-density-800} shows the baryon density in the unit of
normal nuclear density as a function of $\mu_B$.
We read the DCDW onset density as $\rho_B^c\sim 4.7\rho_0$.
This value is larger than the one in Ref.~\cite{Heinz:2013hza}.
We would like to stress that the stable DCDW phase, on the other hand,
appears already at $\rho_B \sim 4.8\rho_0$, while the stable phase
appears at $\rho_B \sim 10.4\rho_0$ in Ref.~\cite{Heinz:2013hza}.
This means that the magnitude of density jump from the uniform state to
the DCDW phase is smaller and therefore the strength of the first order
phase transition is weaker in our case. 
We think that one of the reasons is that the chiral invariant mass is
independent of $\mu_B$, while it depends in the model used in
Ref.~\cite{Heinz:2013hza}.

\begin{figure}[tbp]
 \includegraphics[width=7.cm,clip]{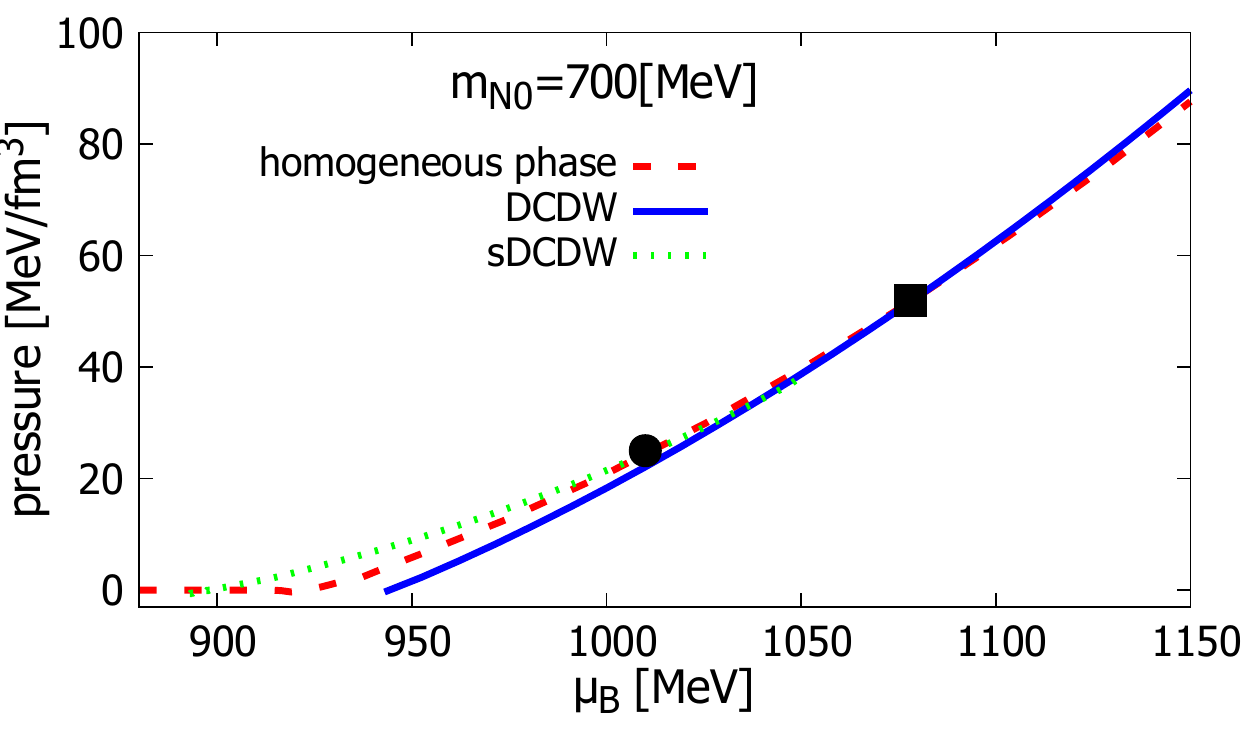}
 \caption{
 Relation between chemical potential and pressure for $m_{0}=700$~MeV.
 The blue solid curve and red dashed curve show the DCDW phase and the
 homogeneous phase, respectively.
 The green dotted curve shows the sDCDW which is another solution with
 $f\neq0$.
 The point of phase transition from the homogeneous phase to DCDW phase
 is expressed by black square, while that from the 
 sDCDW phase to the homogeneous phase is indicated by black circle.
 }
 \label{fig:muB-pres-700-a}
\end{figure}
\begin{figure}[tbp]
\includegraphics[width=7.cm]{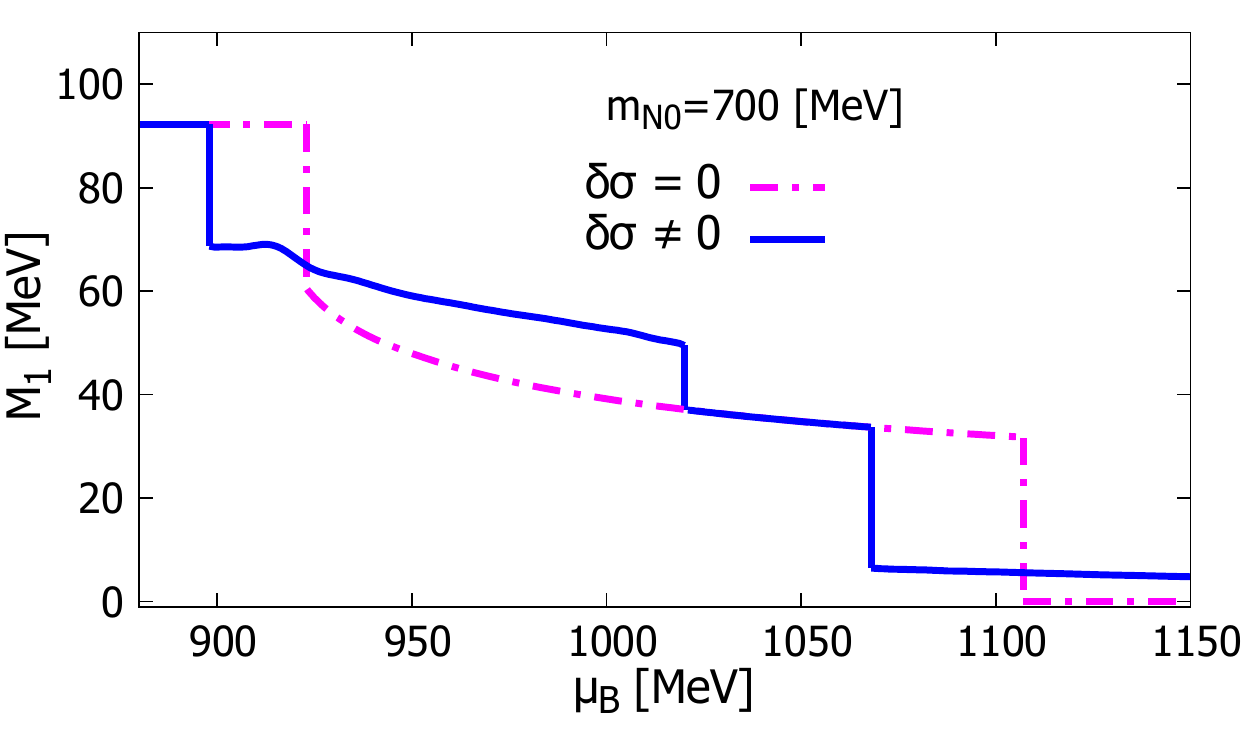}
\caption{
 Relation between chemical potential and $M_1$ for $m_{0}=700$~MeV.
 The blue solid curve and magenta dot-dashed curve show the
 configuration of the extended ansatz with $\delta \sigma \neq 0$ and
 the ordinary ansatz with $\delta\sigma =0$, respectively.
 }
 \label{fig:muB-pres-700-b}
\end{figure}
\begin{figure}[tbp]
 \includegraphics[width=7.cm,clip]{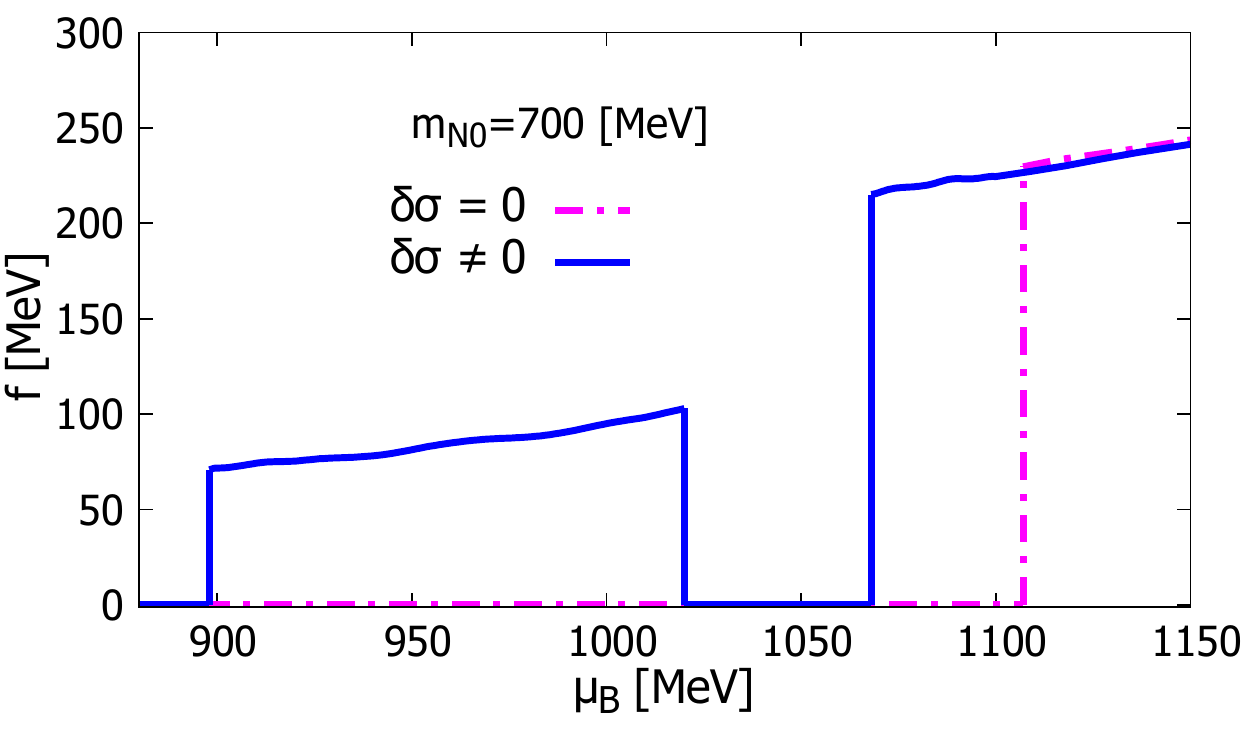}
 \caption{
 Relation between chemical potential and wavenumber $f$ for
 $m_{0}=700$~MeV.
 The blue solid curve shows the solution under the
 extended ansatz with $\delta\sigma\ne 0$, while the magenta dot-dashed
 curve corresponds to the solution with the ordinary ansatz,
 $\delta\sigma=0$.
 }
 \label{fig:muB-pres-700-c}
\end{figure}

\paragraph{$m_{0}=700$~MeV.}
Next, we show the $\mu_B$ dependence of the pressure for $m_{0}=700$~MeV
in Fig.~\ref{fig:muB-pres-700-a}.
We first notice that in this case the normal nuclear matter exists only
as a metastable state.
This is due to the emergence and the stabilization of a new DCDW
state:~%
in addition to the ordinary DCDW phase, we find another solution with
$f\neq0$ which is shown by dotted curve in
Fig.~\ref{fig:muB-pres-700-a}.
To distinguish two solutions with $f\neq0$, we call this phase the
shifted DCDW (sDCDW) phase for a reason described shortly.
We plot the $\mu_B$ dependence of $M_1$ defined in Eq.~(\ref{eq:M1}) in
Fig.~\ref{fig:muB-pres-700-b}, and that of $f$ in the DCDW phase in
Fig.~\ref{fig:muB-pres-700-c}.

From the former, we see that, going up in density, the chiral symmetry
restores via several steps.
From the latter, we clearly see that  there are two regions of DCDW
phase: the sDCDW state with $f \sim 50\,$--$\,100$~MeV for $900\lesssim
\mu_B \lesssim 1020$~MeV, and the ordinary DCDW state with $f
\gtrsim220$~MeV for $\mu_B \gtrsim 1070$~MeV.
Figure~\ref{fig:muB-pres-700-b} shows that the value of $M_1$ for
$\mu_B\gtrsim 1070$~MeV is less than $10$~MeV which is close to the one
in the DCDW phase shown in Fig.~\ref{fig:muB-pres-800-b}, and so is the
value of $f$ ($f \gtrsim220$~MeV).
In fact, this phase is smoothly connected to the ordinary DCDW phase
realized for $m_{0} = 800$~MeV as is seen from the phase diagram
Fig.~\ref{fig:muB-mN0}.
On the other hand, Fig.~\ref{fig:muB-pres-700-c} shows that the value of
$f$ ($f\sim 50\,$--$\,100$~MeV) in the sDCDW phase is less than half of
the value in the ordinary DCDW phase shown in
Fig.~\ref{fig:muB-pres-800-c}.
Furthermore, Figure~\ref{fig:muB-pres-700-b} shows that the value of
$M_1$ for $900 \lesssim \mu_B \lesssim 1020$~MeV is $M_1 \sim
50\,$--$\,70$~MeV which is much larger than $10$~MeV in the ordinary DCDW
phase shown in Fig.~\ref{fig:muB-pres-700-b}.
Since $M_1 = \delta \sigma$ in the DCDW phase, the large difference of
$M_1$ is caused by the difference of the values of $\delta \sigma$:
$\delta \sigma \lesssim 10$~MeV in the ordinary DCDW phase realized for
$\mu_B \gtrsim 1070$~MeV, while $\delta \sigma \sim 50$~MeV for $900
\lesssim \mu_B \lesssim 1020$~MeV.
This implies that the center of the chiral spiral in the
$(\sigma,\pi^0)$ chiral plane, which is near origin in the ordinary DCDW
phase, is shifted to the $\sigma$ direction.
That is why we called the phase realized for $900 \lesssim \mu_B
\lesssim 1020$~MeV the shifted-DCDW (sDCDW) phase.
We would like to stress that, on the contrary to the ordinary DCDW
phase, the solution of stationary condition for $\delta\sigma$ do not
become zero even if it is in chiral limit.

\begin{figure}[tbp]
\begin{center}
\vspace{-80mm}
\includegraphics[bb=0 0 480 600,width=10cm,clip]{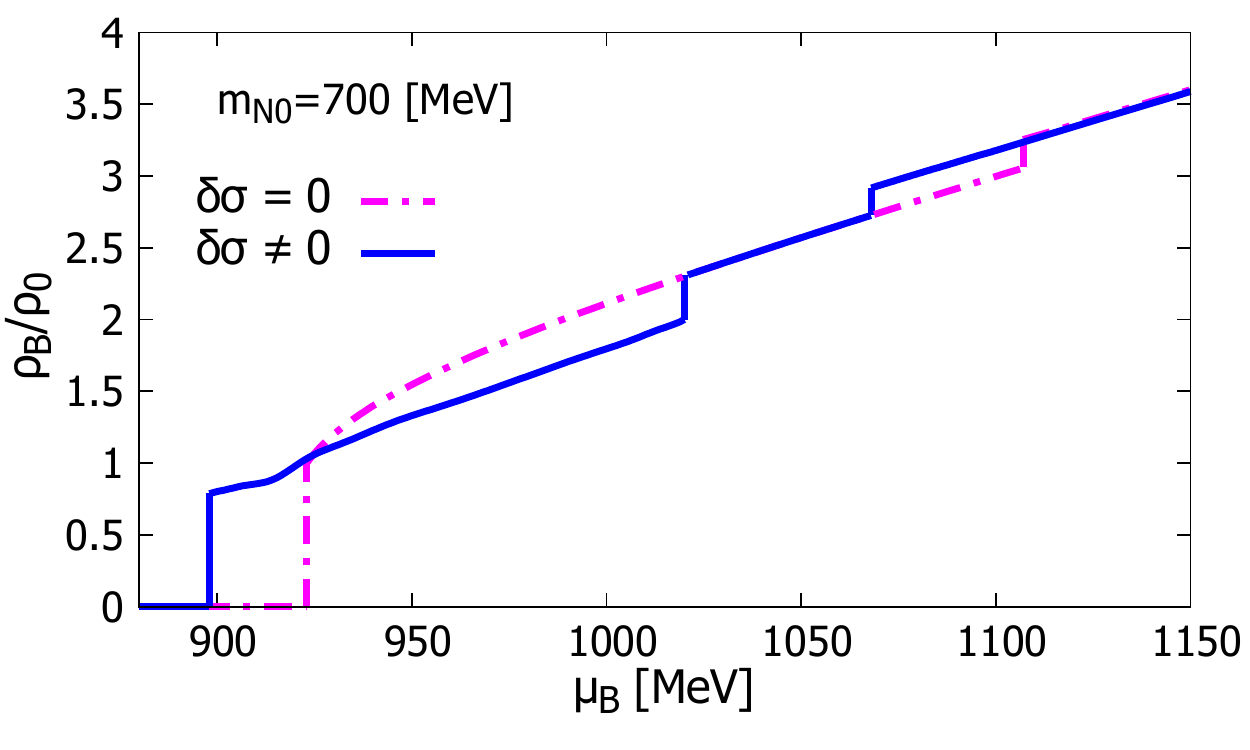}
\caption{
 Relation between chemical potential and baryon number density for
 $m_{0}=700$~MeV.
 The blue solid curve and magenta dot-dashed curve show the
 configuration of the extended ansatz with $\delta\sigma \neq0$ and the
 ordinary ansatz with $\delta\sigma=0$, respectively.
}
\label{fig:muB-density-700}
\end{center}
\end{figure}

The relation between chemical potential and baryon number density is
shown in Fig~\ref{fig:muB-density-700}.
From the figure, we see that the QCD phase next to the vacuum is the
sDCDW phase for $m_{0}=700$~MeV.

\section{A summary and discussions}\label{sec:summary}
We studied the inhomogeneous phase structure in nuclear matter using a
nucleon-based model with parity doublet structure where  $N^\ast(1535)$
is introduced as the chiral partner of $N(939)$.
Adopting the extended ansatz, Eq.~(\ref{eq:ansatz}), we studied the
effect of $\delta\sigma$ and found that, depending on the value of the
chiral invariant mass $m_{0}$, the sDCDW phase exists in addition to the
ordinary DCDW phase.

In the ordinary DCDW phase for large $m_{0}$, the space average of
chiral condensate $M_1$, Eq.~(\ref{eq:M1}), becomes less than 10~MeV,
implying that this phase is smoothly connected to the DCDW phase
obtained with the familiar ansatz $\delta\sigma=0$.
For $m_{0}=800$~MeV, the critical density from the homogeneous phase to
DCDW phase is $4.7\rho_0$.
The wave number $f$ has value of $200\sim300$~MeV which is in fair
agreement with the result obtained in Ref.\cite{Heinz:2013hza}.

On the other hand, when chiral invariant mass $m_{0}\lesssim780$~MeV
the sDCDW phase appears at low density.
This phase is characterized by a smaller wave number $f$ and a large
shift of chiral condensate, $\delta\sigma$.
It is noteworthy that it is not the effect of explicit chiral
symmetry breaking but the dynamical symmetry breaking that produces this
large shift of chiral condensate. 
So we expect that this sDCDW phase survives in the chiral limit.

The parameter range of chiral invariant mass where the sDCDW is
stabilized, fails to realize nuclear matter as the pressure of sDCDW is
so strong that it diminishes the liquid-gas
phase transition structure.
Then, one might think that the present model for $m_{0}$ less than
780~MeV is ruled out.
However, the chiral invariant mass $m_{0}$ can have density dependence
as in Ref.~\cite{Heinz:2013hza} which shows that $m_{0}$ decreases
against increasing density.  In such a case, the sDCDW phase may be
realized in high density nuclear matter in the real world.

Exploring the elementary excitations in the sDCDW phase deserves further
investigations in future.
In the ordinary DCDW phase, apparently both the chiral symmetry and the
translational invariance along $z$-direction are spontaneously broken,
but a particular combination of them is left invariant
\cite{Lee:2015bva,Hidaka:2015xza}.
As a result, the number of spontaneously broken generator is three which
implies that no extra Nambu-Goldstone boson appears other than three
pions.
In contrast, the combination is also spontaneously broken in the sDCDW
phase.
Then, we expect that a phonon mode appears in the sDCDW phase which may
signal the phase.
On the other hand, heavy hadrons 
may also serve as some interesting hard probes in the sDCDW background
\cite{Suenaga:2015daa}.

Another interesting extension of current work is to include an external
magnetic field.
For quark matter, several studies were already devoted to the topic of
inhomogeneous phases under the magnetic field
\cite{Frolov:2010wn,Nishiyama:2015fba,Yoshiike:2015tha,%
Cao:2016fby,Yoshiike:2015wud,Abuki:2016zpv}.
In \cite{Nishiyama:2015fba}, a new DCDW phase was found to occupy the
low density region in an arbitrary small magnetic background.
They called the phase ``weak'' DCDW since it has a smaller value of $f$.
Since they did not  
consider the possible shift of chiral condensate,
$\delta\sigma$, the analysis within the nucleon-based model with
including both $\delta\sigma$ and magnetic field would be an interesting
subject worth exploring.

\smallskip
\noindent
{\it Acknowledgement.}~This work was partially supported by JPSP KAKENHI
Grant Number JP16K05346 (H.A.) and 16K05345 (M.H.).

\end{document}